\documentclass[prd,aps,tightenlines,twocolumn]{revtex4}
\usepackage{graphicx,color}
\usepackage{amssymb}

\begin{document}
\title{Graviton emission from a Gauss-Bonnet brane}
\author{Kenichiro Konya}
\affiliation{Institute for Cosmic Ray Research, 
University of Tokyo, Kashiwa 277-8582, Japan}

\date{\today}
\begin{abstract}
We study the emission of gravitons by a homogeneous brane 
with the Gauss-Bonnet term into  an Anti de Sitter five dimensional bulk spacetime.
It is found that the graviton emission depends on the curvature scale and 
the Gauss-Bonnnet coupling  and that the amount of emission generally decreases.  
Therefore nucleosynthesis constraints  are easier to satisfy by including the Gauss-Bonnet term.
\end{abstract}

\maketitle

\section{Introduction}
In recent years there has been considerable interest in the suggestion that 
our universe is a brane: a sub-space embedded in a higher-dimensional bulk spacetime. 
In these models, ordinary matter is confined to our brane while the gravitational field 
propagates through the whole spacetime.  Of particular importance is the Randall-Sundrum 
(RS) model, where a single brane is embedded in an infinitely extended  Ad$S_{5}$ spacetime  
\cite{RS2}.  At low energies, the zero mode of the 5D graviton is localized on the brane, 
because of the strong curvature of the bulk due to a negative bulk cosmological constant, 
and 4D gravity is recovered.  \\
~A natural extension of the RS model is to include higher order curvature invariants 
in the bulk. Such terms arise in the Ads/CFT correspondence as next-to-leading order 
corrections to the CFT \cite{CFT}. Particularly, in the heterotic string effective action, 
the Gauss-Bonnet (GB) term arises as the leading order quantum corrections to gravity. 
This gives the most general action with second-order field equations in five dimensions 
\cite{Lovelock,Lanczos} and is investigated in areas such as black hole\cite{GBBH}
 and brane-world.
The graviton is localized in the GB 
 brane-world \cite{GBlocal} and deviations from Newton's law at low energies are less 
 pronounced than in the RS model \cite{GBnew}. \\
 ~Brane cosmologies with and without GB term has been investigated \cite{GBcosmo,RSrad1,RSrad2}. 
 Due to the cosmological symmetries, most GB brane-world scenarios assume that the 5D spacetime 
 metric is the generalized Schwartzschild-Anti de Sitter (Sch-Ads), described by the metric \cite{sp,GB}
\begin{eqnarray}
ds^{2} = -g(r)dt^{2} + \frac{dr^{2}}{g(r)} +r^{2}\Omega_{ij}dx^{i}dx^{j}, \nonumber \\
g(r) = k + \frac{r^{2}}{4\alpha}\left(1-\sqrt{1+\frac{4}{3}\alpha\Lambda+8\alpha\frac{C}{r^{4}}}\right),
\end{eqnarray}
where $\Omega_{ij}$ is the three dimensional metric of space with constant curvature $k = -1,0,1$, 
$\Lambda$ is the bulk cosmological constant, $\alpha > 0$ represents the GB coupling.
In the $\alpha \rightarrow 0$ limit, this reduces to the usual Sch-Ads metric.  If the bulk is empty then $C$ is 
necessarily constant in time. However particle interactions can produce gravitons that are emitted  
into the bulk at high energies on a brane. Therefore, in a realistic cosmological scenario, there 
exists an avoidable bulk component and so $C$ is no longer constant. This problem has recently 
been  studied for a RS brane \cite{RSrad1,RSrad2}. In this paper we examine the radiating  GB brane-world and 
find what effects including the GB term has on the evolution of $C$.\\
  ~~The rest of this paper is organized as follows: in section II  we derive the energy loss through 
  graviton radiation; in section III we derive the emission rate of the bulk gravitons; in section IV we numerically solve the system of equations under some 
  approximations; in section V some conclusions are drawn.

\section{The bulk and the brane} 
In order to model the bulk spacetime metric, we use the five dimensional and Gauss-Bonnet 
generalization of the Vaidya metric given by \cite{GBVaidya},
\begin{eqnarray}
ds^{2} = -f(r,v)dv^{2} +2drdv +r^{2}\Omega_{ij}dx^{i}dx^{j}, \nonumber \\
f(r,v) = k + \frac{r^{2}}{4\alpha}\left(1-\sqrt{1+\frac{4}{3}\alpha\Lambda+8\alpha\frac{C(v)}{r^{4}}}\right)
\end{eqnarray}
where $v = $ const are ingoing plane-formed null rays. If $C$ does not depend on $v$, then the metric 
(2) is a rewriting of the generalized Sch-Ads metric (1), as can be seen by the coordinate transformation 
$v = t+ \int dr/f(r)$. From now on we assume that the brane is outside the horizon ($f > 0$) and that the brane 
universe is spatially flat.  The Vaidya type metric is a solution to  Einstein-Gauss-Bonnet equations 
\begin{eqnarray}
G_{ab} + 2\alpha H_{ab} + \Lambda g_{ab} = \kappa^{2}T_{ab},
\end{eqnarray}
where
\begin{eqnarray}
G_{ab} = R_{ab} - \frac{1}{2}g_{ab}R,~~~~~~~~~~~~~~~~~~~~~\\
H_{ab} = RR_{ab} -2R_{ac}R^{c}_{~b} -2R^{cd}R_{acbd}~\nonumber\\
+R_{a}^{~cde}R_{bcde} -\frac{1}{4}g_{ab}L_{GB},~~~~~~
\end{eqnarray}
and the bulk energy-momentum tensor has null-radiation form,
\begin{eqnarray}
T_{ab} = \psi k_{a}k_{b}.
\end{eqnarray}
Here, $\kappa^{2} \equiv 1/M^3$ is  
the five dimensional gravitational coupling, $\psi$ is, for a brane observer, 
the flux of gravitons leaving a radiation dominated brane and
$k_{a}$ is a null vector. Thus, in our model the bulk gravitons are presumed to be emitted radially. 
By inserting the metric (2) and the stress energy tensor (6) into 
 Einstein-Gauss-Bonnet equations (3) we find the evolution equation for $C$:
\begin{eqnarray}
\frac{dC}{dv} = \frac{2\kappa^{2}\psi r^{3}}{3}k_{v}k_{v}.
\end{eqnarray}
The appropriate normalization of $k_{a}$ is given by $k_{a}u^{a} = -1$, where $u^{a}$ is the brane's 
velocity vector. This implies that the only nonvanishing component is $k^{r} = k_{v} = 1/\dot{v}$, 
where $\dot{v} = dv/d\tau$ and $\tau$ is cosmic proper time on the brane.
  From $u^{a}u_{a} = 1$ we obtain 
\begin{eqnarray}
f\dot{v} = \dot{r} + \sqrt{\dot{r}^{2} + f}.
\end{eqnarray}

In order to determine the behavior of the brane we have to impose the generalized Israel junction 
conditions which are given by \cite{gIsrael},
\begin{eqnarray}
[K_{\mu\nu}] -h_{\mu\nu}[K] + 2\alpha(3[J_{\mu\nu}]-h_{\mu\nu}[J]
-2P_{\mu\rho\nu\sigma}[K^{\rho\sigma}])  \nonumber\\ 
= -\kappa^{2}S_{\mu\nu},
\end{eqnarray}
where 
\begin{eqnarray}
J_{ab} = \frac{1}{3}(2KK_{ac}K^{c}_{~b} + K_{cd}K^{cd}K_{ab} ~~~~~~\nonumber\\
-2K_{ac}K^{cd}K_{db}-K^{2}K_{ab}), \\
P_{\mu\nu\rho\sigma} = R_{\mu\nu\rho\sigma} +2h_{\mu[\sigma}R_{\rho]\nu}
+ 2h_{\nu[\rho}R_{\sigma]\mu} \nonumber\\
 + Rh_{\mu[\rho}h_{\sigma]\nu}.~~~~~~~~~~ ~~~~~~~~~~~~~~~
\end{eqnarray}
Here, $K_{ab}$ is the extrinsic curvature, $h_{\mu\nu}$ is the induced metric on the brane
 and $S_{\mu\nu}$ is the brane energy momentum tensor. From these junction conditions 
 we obtain the following Friedmann equation \cite{sp,gIsrael}:
 \begin{eqnarray}
H^{2} = \frac{c_{+} + c_{-}-2}{8\alpha},
\end{eqnarray}
where 
\begin{eqnarray}
c_{\pm} =  \left\{\left[\left(1+\frac{4}{3}\alpha\Lambda + \frac{8\alpha C}{r^4}\right)^{3/2} 
+ \frac{\alpha}{2}\kappa^{4}\sigma^{2}\right]^{1/2} \right.  \nonumber\\
\left.\pm\sqrt{\frac{\alpha}{2}}\kappa^{2}\sigma\right\}^{2/3},~~~~~~~~~~~~~~~~~~~~~~~~~~~~~
\end{eqnarray}
and $\sigma$ represents the energy density of the matter source.  The requirement 
that the standard form of the Friedmann equation is recovered at sufficiently low energy 
scales  leads to the relation 
\begin{eqnarray}
\kappa_4^2\equiv \frac{1}{M_{Pl}^2} = \frac{\kappa^2}{(1+\gamma)\ell}, 
\end{eqnarray} 
where $M_{Pl}$ is the reduced 4D Planck scale, 
$\ell^{-2} = (1-\sqrt{1+4\alpha\Lambda/3})/4\alpha$ is the AdS curvature scale, 
$\gamma = 4\alpha/\ell^2$, and 
we have the standard assumption that the energy density on the brane is separated two parts, 
the ordinary matter component, $\rho$, and the brane tension, $\lambda$, such that $\sigma = \rho + \lambda.$ We also assume zero cosmological constant on the brane,
\begin{eqnarray}
\kappa^2\lambda =\frac{2(3-\gamma)}{\ell}. 
\end{eqnarray}
The GB term is considered as the lowest-order stringy correction to the 5D Einstein action, so the GB energy scale should be larger than the RS energy scale. 
From this consideration, we have $\gamma \leq 0.15$ \cite{GBRS}. \\
~~Then the Raychaudhuri equation is written as 
\begin{eqnarray}
\dot{H} + H^2 = -\frac{1-b^{1/3}}{4\alpha}-\frac{2C}{r^4b^{1/3}} - \frac{\kappa^2\psi}{3b^{1/3}}
~~~~~~~~~~~~~ \nonumber \\
-\left\{\left[2\left(b^{1/3}-\frac{8\alpha C}{r^4b^{1/3}} -\frac{4\kappa^2\alpha\psi}{3b^{1/3}}\right)
\left(H^2+\frac{1-b^{1/3}}{4\alpha}\right) \right. \right. \nonumber\\
\left. +\frac{\kappa^2}{3}(\rho-3\lambda)\sqrt{H^2+\frac{1-b^{1/3}}{4\alpha}}\right]  \nonumber \\
\left./ \left[1+8\alpha\left(H^2+\frac{1-b^{1/3}}{8\alpha}\right)\right]\right\},
\end{eqnarray}
 where $b^{1/3} = \sqrt{1+4\alpha\Lambda/3+8\alpha C/r^4}$. \\
~~Because of graviton emission, the brane energy is not conserved,
\begin{eqnarray}
\dot{\rho} +4H\rho = -2\psi,
\end{eqnarray}
The factor of 2 on the right hand side is due to the fact that the brane is radiating a flux of gravitons 
into both sides.

\section{Production rate of bulk gravitons}
In order to determine quantitatively the energy loss $\psi$ we follow the same procedure 
 as in the RS case \cite{RSrad1}.
First, we evaluate  the cross section of the process $\phi + \bar{\phi} \rightarrow$ KK graviton, 
where $\phi$ is a particle confined on the brane. To compute this cross section we have to
check  whether or not the cosmological influence is negligible.  In the GB high energy regime 
the Hubble rate is approximated as $H \sim T^{4/3}/\alpha^{1/3}M$, where $T$ is the temperature  
of the brane particle. Here,  the GB energy scale should be smaller than $M$ so that there 
is the GB regime 
before the quantum gravity regime. From this condition we have $\alpha \gg M^{-2}$ \cite{GBRS}. 
Therefore, the temperature $T$ is bigger than the Hubble rate $H$ in the GB regime if we assume $T \ll M$. 
 Even after the GB regime $T \gg H$ as shown in \cite{RSrad1}. Thus, we find that the cosmological
  influence can always be neglected. \\ 
 ~~ Let us consider the linear perturbations of the GB metric,  
 \begin{eqnarray}
ds^2 = e^{-2A(z)}\left\{ (\eta_{\mu\nu}+h_{\mu\nu})dx^\mu dx^\nu +dz^2   \right\},
\end{eqnarray}
in axial gauge with 4d TT condition.  
Here, $A(z) = \log(|z|/\ell+1)$. 
We decompose the graviton into KK modes, 
\begin{eqnarray}
h_{\mu\nu} = \int dm u_{m}(z)\phi_{\mu\nu}(x),
\end{eqnarray}
where the modes $u_m(z)$ are given by  \cite{w-function},
\begin{eqnarray}
u_m(z) = \sqrt{\frac{m(|z|+\ell)}{2(1+A^2)}}\left[Y_2(m(|z|+\ell)) \right. \nonumber \\
\left. +AJ_2(m(|z|+\ell))  \right], \\
A = -\frac{Y_1(m\ell)+\chi m\ell Y_2(m\ell)}{J_1(m\ell)+\chi m\ell J_2(m\ell)},~~~~~
\end{eqnarray}
and satisfy the normalization $\int dz u_m^*(z)u_{m'}(z) = \delta(m-m')$. 
$Y$ and $J$ are the Bessel functions and Neumann functions respectively, and 
$\chi = \gamma/(1-3\gamma)$.
In the $\alpha \rightarrow 0$ limit we recover the RS result. 
The coupling of the bulk graviton to the brane matter is described by the action
\begin{eqnarray}
S_{int} = \kappa(1-\gamma)^{1/2}\int dm u_{m}(0)\int d^4xS^{\mu\nu}\phi_{\mu\nu}. 
\end{eqnarray}
The overall factor $(1-\gamma)^{1/2}$ is a  GB correction \cite{GBnew,w-function}. 
From this action we can calculate the amplitude for the scattering of brane particles 
leading to a KK emission. This calculation is quite analogous to the procedure 
already described in the context of flat extra dimensions \cite{ampli}, the only difference 
being the coupling constant in (22). Using those results, one finds that the spin and particle-anti 
particle averaged squared amplitude is given by 
\begin{eqnarray}
\Sigma |{\cal M}|^2 = \kappa^2(1-\gamma)|u_m(0)|^2A\frac{s^2}{8},
\end{eqnarray}
where $s= (p_1+p_2)^2$ ($p_1$ and $p_2$ being the incoming four-momenta of the scattering particles), 
and 
\begin{eqnarray}
A = \frac{2}{3}g_s+g_f+4g_v
\end{eqnarray}
where $g_s$, $g_f$ and $g_v$ are respectively the scalar, fermion, and vector relativistic degrees of freedom.  
To derive this amplitude we assume that the mass of the incoming particles is neglected. \\ 
~~Going back to cosmology, the production of gravitons results into an energy loss for ordinary matter, 
which can be expressed as 
\begin{eqnarray}
\dot{\rho} + 4H\rho = -\int dm\int\frac{d^3p_m}{(2\pi)^3}\mathbf{C}[f],
\end{eqnarray}
 with 
\begin{eqnarray}
\mathbf{C}[f] = \frac{1}{2}\int\frac{d^3p_1}{(2\pi)^32E_1}\int\frac{d^3p_2}{(2\pi)^32E_2} \nonumber \\
\times\Sigma|{\cal M}|^2f_1f_2(2\pi)^4\delta^{(4)}(p_1+p_2-p_m)
\end{eqnarray}
where $f_i = 1/(e^{E_i/T}\pm1)$ is the Fermi/Bose distribution function and $p_m$ is the 
four-momentum of the created bulk graviton. The graviton production can be significant at high energies. 
So, heavy gravitons with $m \sim T \gg \ell^{-1}$ mainly contribute to the energy loss (this 
 is a good approximation for a range of values for $\ell$ since the constraint from gravity experiment 
 is  $\ell < 10^3~\textmd{eV}^{-1}$). Accordingly we have to look into the behavior of the 
 mode function $u_m$ for $m \gg\ell^{-1}$. In the RS case we have $|u_m(0)|^2 =
 $ const for $m \gg \ell^{-1}$. However, in the 
 GB case the mode functions exhibit a rather nontrivial dependence on $m$ for $m \gg \ell^{-1}$ as 
 shown Figure 1.  For the modes $m\gg\ell^{-1}$,  eqs. (20) and (21) give us 
 \begin{eqnarray}
u_m(0) \simeq -\sqrt{\frac{1}{\pi}}\frac{1}{\sqrt{1+3\chi+\frac{15}{8}\chi^2+\chi^2m^2\ell^2}},
\end{eqnarray}
 where we have neglected the term which is smaller than $m^{-2}\ell^{-2}$.
 
\begin{figure}[t]
\begin{center}
\includegraphics[width=8.5cm]{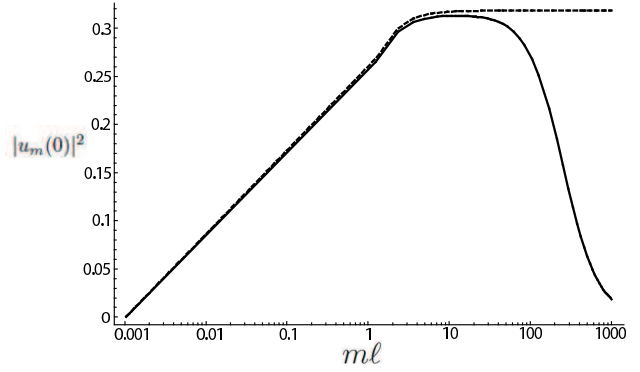}
\caption{The graviton mode functions evaluated on the brane as a function of Kaluza-Klein mass $m$. 
The solid line is the function including  the GB term, and the doted line is the one without GB term. 
The adopted parameter is $\chi = 0.004$.}
\end{center}
\end{figure}

 From eqs. (17), (23), (25), (26) and (27) we find that 
\begin{eqnarray}
\psi \simeq \frac{\kappa^2A\left(1-\gamma\right)}{2^9\pi^3}\sum_{n=1}^{\infty}\frac{2nT^8}{1+n}\times~~~~~~~~~~~~~~~~~~~~~~~~~
 \nonumber\\
\int dx\frac{x^6K_2\left(\frac{1+n}{2}x\right)}{\pi\left(1+3\chi+\frac{15}{8}\chi^2+\chi^2\ell^2T^2x^2\right)}
~~ \nonumber  \\
\sim \left\{ \begin{array}{c}
\frac{\alpha_{GB}(1-\gamma)\kappa^2\rho^{3/2}}{\chi^2\ell^2}~~~~~~~
\textmd{for}~m\sim T \gg 1/\chi\ell \\
\frac{\alpha_{RS}(1-\gamma)\kappa^2\rho^{2}}{1+3\chi+\frac{15}{8}\chi^2}
~\textmd{for}~ \ell^{-1}\ll m\sim T \ll 1/\chi\ell
\end{array},\right.
\end{eqnarray}
 where $K_2$ is the modified Bessel function of the second kind, and $\alpha_{GB,RS}$ is 
 a dimensionless  constant related to the total number of relativistic degrees of freedom. If all 
 degrees of freedom of the standard model are relativistic, $\alpha_{GB}\simeq 4.56\times 10^{-4}$ 
 and $\alpha_{RS}\simeq 1.54\times 10^{-3}$.  We find  that $\psi$ is proportional to $\rho^{3/2}$, 
 not $\rho^2$ as  in the RS case at high energy scales \cite{GBVaidya}.  Note that the transition energy 
 scale depends on the bulk curvature scale $\ell$ and that this energy scale is  lower than the 
 RS energy scale for a wide range of values of $\ell$ and $\alpha$.

 \section{Numerical analysis}
It is useful to define the dimensionless parameters, $\hat{\rho} = \rho/\lambda$, 
$\hat{t} = t/\ell$, $\hat{H} = H/\ell$, $\hat{C} = C/\ell^2$, and $\hat{\alpha} = \alpha/\ell^2$. 
The first four variables are the same as those used in Refs. \cite{RSrad1,RSrad2}. 
Using these variables, the dynamics on the brane are governed by the following system of 
differential equations:
\begin{eqnarray}
\frac{d\hat{\rho}}{d\hat{t}} + 4\hat{H}\hat{\rho} = -\hat{\psi}, ~~~~~~~~~~~~~~~~~~~~~~~~\\
\frac{d\hat{C}}{d\hat{t}} = \frac{2}{3}(3-\gamma)\hat{\psi}r^4\left(\sqrt{\hat{H}^2+\frac{1-b^{1/3}}{4\hat{\alpha}}}
-\hat{H}\right), ~~~~\\
\frac{d\hat{H}}{d\hat{t}} = -\hat{H}^2 -\frac{1-b^{1/3}}{4\hat{\alpha}} -\frac{2\hat{C}}{r^4b^{1/3}} 
-\frac{(3-\gamma)\hat{\psi}}{3b^{1/3}}~~~~~ \nonumber \\
-\left\{2\left(b^{1/3}-\frac{8\hat{\alpha}\hat{C}}{r^4b^{1/3}}-\frac{\gamma(3-\gamma)\hat{\psi}}{3b^{1/3}}\right)\left(\hat{H}^2+\frac{1-b^{1/3}}{4\hat{\alpha}}\right) \right. \nonumber\\
\left.+(1-\frac{\gamma}{3})
\left(\hat{\rho}-3\right)\sqrt{\hat{H}^2+\frac{1-b^{1/3}}{4\hat{\alpha}}} \right\} \nonumber \\
/\left\{1+8\hat{\alpha}\left(\hat{H}^2+\frac{1-b^{1/3}}{8\hat{\alpha}}\right)\right\},~~
\end{eqnarray}
where 
 \begin{eqnarray}
\hat{\psi} = \frac{2\ell}{\lambda}\psi ~~~~~~~~~~~~~~~~~~~~~~~~~~~~~~~~~~~~~~~
~~~~~~~~~~~~~~~~~~~~~~\nonumber \\
\sim \left\{ \begin{array}{c}
2\alpha_{GB}(1-\gamma)\sqrt{\frac{2(3-\gamma)}{1+\gamma}}\left(\frac{\kappa_4}{\ell}\right)
\frac{\hat{\rho}^{3/2}}{\chi^2}~~\textmd{for}~ T^2 \gg 1/\chi^{2}\ell^{2}\\
\frac{4\alpha_{RS}(3-\gamma)(1-\gamma)\hat{\rho}^{2}}{1+3\chi+\frac{15}{8}\chi^2}
~~~~\textmd{for}~ \ell^{-2}\ll T^2 \ll 1/\chi^{2}\ell^{2}~~~~~~~~
\end{array}.\right.
\end{eqnarray}
In order to derive above equations we use eqs. (7), (14), (15), (16), (17) and (28).  
Unfortunately, it is very difficult to find a general analytic solution to these equations, 
such as that found by Leeper et. al. for the RS case \cite{RSradsolution}. So, we use the 
approximation  of eq. (32) and solve the above coupled system numerically. 
Here, the algebraic constraint from the generalized Friedmann equation,
\begin{eqnarray}
\hat{H}^2 = \frac{\hat{c}_++\hat{c}_--2}{8\hat{\alpha}},~~~~~~~~~ \nonumber \\
c_{\pm} = \left\{\left[b+2(3-\gamma)^2\hat{\alpha}\left(1+\hat{\rho}\right)^2 \right. \right. \nonumber \\
\left.\left.\pm\sqrt{2\hat{\alpha}}(3-\gamma)(1+\hat{\rho})
\right]\right\}^{2/3}
\end{eqnarray}
is used to monitor numerical errors. Results from a numerical integration of this system 
with a variety of initial conditions for $\alpha$ and $\ell$ are shown in Figures 2-5.  
The initial value of $\hat{\rho}_i$ is taken $10^4$ except that $10^4$ is larger than the 
highest energy density scale $M^4/\lambda$ \footnote{We confirm that there are no big differences
if we take $\hat{\rho}_i = M^4/\lambda$.}.  In such cases we take $\hat{\rho}_i = M^4/\lambda$.

\begin{figure}[t]
\begin{center}
\includegraphics[width=8.5cm]{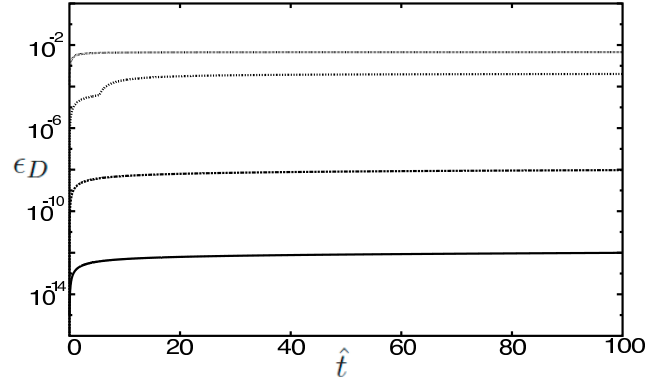}
\caption{Evolution of $\epsilon_D$ with $\ell =1\textmd{GeV}^{-1}$. The curves from top 
to bottom correspond to the cases $\hat{\alpha} = 10^{-10}, 10^{-9}, 10^{-7} 
\textmd{and} ~10^{-5}$. 
We can see that the transition between $\psi\propto\rho^{3/2}$ and $\psi\propto\rho^{2}$ occurs in the 
second curve from above.  }
\label{+branch}
\end{center}
\end{figure}

Figure 2 shows the effect of  increasing $\alpha$ while keeing $\ell$ fixed. 
Here, we define $\epsilon_D$ as the ratio of dark radiation to standard radiation energy density:
\begin{eqnarray}
\epsilon_D \equiv \frac{9\hat{C}}{2(3-\gamma)^2r^4\hat{\rho}}. 
\end{eqnarray} 
The first thing to notice is that the larger $\alpha$ is the smaller $\epsilon_D$. 
This is because the interaction between brane matter and the bulk gravitons weakens due 
to the GB term as can be seen in Figure 1 and  the brane emits less gravitons. \\
~The evolution of $\epsilon_D$ with $\hat{\alpha} = 10^{-7}$ is shown in Figure 3. 
There is also a marked effect on $\epsilon_D$.  The increase of  $\ell$ leads to 
an extension of the $\psi \propto \rho^{3/2}$ regime and a suppression of  graviton emission  
by  eq. (28).

\begin{figure}[t]
\begin{center}
\includegraphics[width=8.5cm]{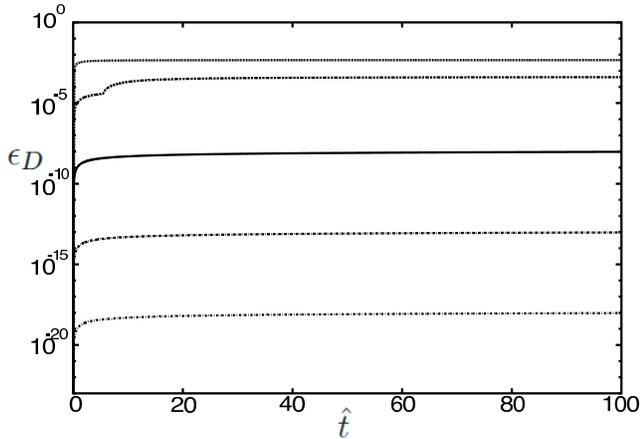}
\caption{Evolution of $\epsilon_D$ with $\hat{\alpha} =10^{-7}$. The curves from top 
to bottom correspond to the cases $\ell = 10^{-7}, 10^{-5}, 1, 10^{-5}, \textmd{and} ~10^{10}\textmd{GeV}^{-1}$. 
We can also see that the transition between $\psi\propto\rho^{3/2}$ 
and $\psi\propto\rho^{2}$ occurs in the 
second curve from above. }
\end{center}
\end{figure}

At low energies, dark radiation is produced at a negligible rate so that there is an asymptotic 
constant value for $\epsilon_D$ as shown in Figures 2 and 3. These asymptotic  values of $\epsilon _D$ 
are shown in Figures 4 and 5. We find that there is an upper bound on $\epsilon_D$ and that 
this upper bound value is the  final value of the RS case \cite{RSrad1}:
\begin{eqnarray}
\epsilon_{D}^{GB} < \epsilon_{D}^{RS} \rightarrow 3\alpha_{RS}.
\end{eqnarray}
This quantity is constrained by cosmological observations. The number of additional 
relativistic degrees of freedom is usually measured in units of extra neutrino species $\Delta N_{\nu}$. 
A typical bound $\Delta N_{\nu} \hspace{0.3em}\raisebox{0.4ex}{$<$}\hspace{-0.75em}\raisebox{-.7ex}{$\sim$}\hspace{0.3em}1$ \cite{extranu}
 implies $\epsilon_D \hspace{0.3em}\raisebox{0.4ex}{$<$}\hspace{-0.75em}\raisebox{-.7ex}{$\sim$}\hspace{0.3em} 0.35$. This bounds  is just above the estimated value 
 for the RS case.  Including the GB term can help reduce the final value of the dark radiation term 
 and hence  we can easily satisfy this bound. \\
\\

\begin{figure}[t]
\begin{center}
\includegraphics[width=8.5cm]{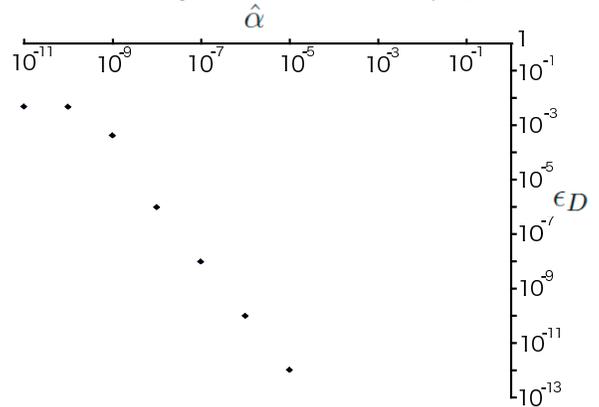}
\caption{The asymptotic values  of $\epsilon_D$ with $\ell = 1\textmd{GeV}^{-1}$.}
\end{center}
\end{figure}

\begin{figure}[t]
\begin{center}
\includegraphics[width=8.5cm]{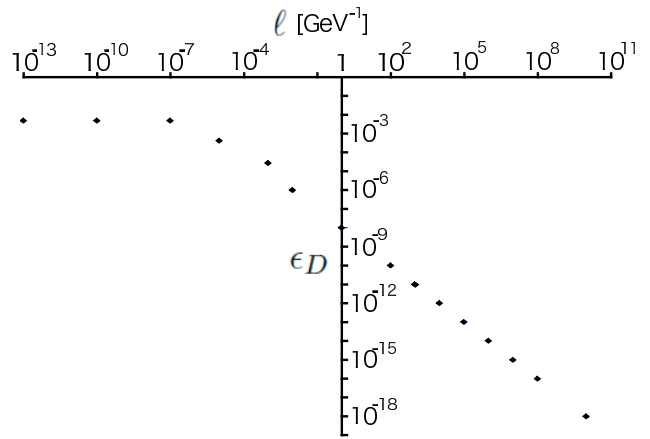}
\caption{The asymptotic values  of $\epsilon_D$ with $\hat{\alpha} = 10^{-7}.$}
\end{center}
\end{figure}

\section{Conclusions}
In this paper, we have considered a GB brane that emits gravitons at early times, 
using a generalized Ads-Vaidya spacetime approximation. We have derived the dynamical equations 
governing the evolution of the energy density $\rho$, the scale factor $r$, and the dark radiation 
parameter $C$ in section II. In section III we have derived the production rate of bulk gravitons. \\ 
~We have performed numerical integration of the system of differential equations in section IV. 
The different feature 
from a RS radiating brane is that the asymptotic value for the dark radiation depends on the 
curvature scale $\ell$ and the GB coupling term $\alpha$. And we have  demonstrated that  the late-time 
dark radiation is generally suppressed and so cosmological limits can be easily satisfied when 
there is a GB term.

\acknowledgments
 We would like to thank M.~Kawasaki  for the helpful advices.

\end{document}